\documentclass[10pt, conference, letterpaper]{IEEEtran}
\usepackage[table]{xcolor}
\usepackage[utf8]{inputenc}
\usepackage{csquotes}
\usepackage{graphicx}
\graphicspath{ {./resources/} }
\usepackage{mathptmx}
\usepackage{authblk}
\newcommand{\ignore}[1]{}
\newcommand{\ignoreA}[1]{}
\usepackage{verbatim}
\usepackage{pdfpages}
\usepackage{lipsum}
\usepackage{amsmath}
\usepackage{varwidth}
\usepackage{caption}
\usepackage{url}
\usepackage{subcaption}
\usepackage{paralist}
\usepackage{tablefootnote}

\usepackage{graphicx}
\title{Technical Report: Performance Comparison of Service Mesh Frameworks: the MTLS Test Case}

\author{ 
    Anat Bremler Barr\textsuperscript{*}, Ofek Lavi\textsuperscript{*}, Yaniv Naor\textsuperscript{\dag}, Sanjeev Rampal\textsuperscript{\ddag}, Jhonatan Tavori\textsuperscript{*} \\
    anatbr@tauex.tau.ac.il, ofeklavi@mail.tau.ac.il, yanivnaor92@gmail.com, \\srampal@redhat.com, jhonatant@mail.tau.ac.il \\
    \textsuperscript{*}Tel Aviv University, \textsuperscript{\dag}Reichman University, \textsuperscript{\ddag}Red Hat
}

\begin{document}

\maketitle

\begin{abstract}
Service Mesh has become essential for modern cloud-native applications by abstracting communication between microservices and providing zero-trust security, observability, and advanced traffic control without requiring code changes. This allows developers to leverage new network capabilities and focus on application logic without managing network complexities. However, the additional layer can significantly impact system performance, latency, and resource consumption, posing challenges for cloud managers and operators.

In this work, we investigate the impact of the mTLS protocol—a common security and authentication mechanism—on application performance within service meshes. Recognizing that security is a primary motivation for deploying a service mesh, we evaluated the performance overhead introduced by leading service meshes: \textit{Istio}, \textit{Istio Ambient}, \textit{Linkerd}, and \textit{Cilium}. Our experiments were conducted by testing their performance in service-to-service communications within a \textit{Kubernetes} cluster.

Our experiments reveal significant performance differences (in terms of latency and memory consumption)  among the service meshes, rooting from the different architecture of the service mesh, sidecar versus sidecareless, and default extra features hidden in the mTLS implementation. 
 Our results highlights the understanding of the service mesh architecture and its impact on performance.

\end{abstract}

    \section{Introduction}
  \label{chap:introduction}

   A service mesh is a dedicated infrastructure
   which controls service-to-service communication over a network and provides a way to control how different parts of an application share data with one another. 
    Service mesh is most commonly used on top of the Kubernetes platform to overcome the challenges of a large-scale microservice-based system.
   A survey of the Cloud Native Computing Foundation (CNCF) \cite{cncf_sm_survey_2022} found that 70\% of the respondents run a service mesh in production or development and 19\% are in evaluation.


    The number of service mesh providers is constantly growing, with each offering different features, performance, ease of use, and pricing, while continuously expanding capabilities. Typically, third parties develop and maintain the service mesh layer, allowing application developers to focus on business logic without worrying about network complexities. As services work by responding to incoming requests and issuing outgoing requests, the flow of requests becomes a critical determining factor of how the application behaves at runtime. Thus, standardizing the management of this traffic is crucial for guaranteeing the application’s runtime.

    A CNCF survey \cite{cncf_sm_survey_2022} examined factors driving organizations to adopt service meshes. Security was in the top concerns, with 79\% of respondents relying on techniques like mTLS to reduce attack risks. This aligns with the growing adoption of the zero-trust security model, which requires all system users to be authenticated before accessing data \cite{zero_trust_architecture}. 
    
    While service meshes offer benefits like improved reliability and consistency, their impact on application performance remains a concern for software architects. 
    The performance overhead introduced by service meshes, typically assessed through metrics like throughput, latency, and resource consumption (CPU and memory), arises from the additional hops required for each system call. This overhead can increase request and response times, potentially reducing the overall benefits that service meshes are designed to offer.
  
    The goal of this work is to compare the performance of four common service mesh models—Istio \cite{istio}, Istio Ambient \cite{Istio-ambient}, Linkerd\cite{linkerd}, and Cilium \cite{cilium}—in terms of latency and resource usage. Further, we aim to study the following core questions: How does the mTLS protocol affect network performance? What is the performance cost of offloading mTLS logic to the service mesh? We seek to offer a comprehensive assessment of the trade-offs involved in using service meshes in performance-critical environments.
  
    To evaluate the service mesh frameworks, we set up a testing environment that simulates a production-like cloud environment.  During testing, we monitor the system's behavior under load through standard monitoring tools such as Prometheus\cite{prometheus}.\ignoreA{This environment will be used for the performance tests. We designed a test suite that should give us insights into the service mesh behavior under load and what are the factors that contribute to it. We begin with a baseline test that is used as a comparison point for future tests and then we test each service mesh separately. During the tests, we monitor the system's behavior through standard monitoring tools such as Prometheus \cite{prometheus}. At the end of each test, we collect and save the results. 
    
    We provide a comparison and analysis of a few of the popular service mesh frameworks: , \emph{Linkerd}, and \emph{Cilium}, focusing on overall performance and the impact of various configurations. The results are analyzed, and compared between all service meshes in order to answer our research questions. We aim to help software developers and management better understand service meshes and assist in selecting the most suitable one for their systems. }

Table \ref{fig:summary-table} summarizes the main difference between the service-mesh architectures different characteristics (explained in detailed in Section \ref{sec:background}) and the main experiment results (explained in detailed in Section \ref{chap:tests-results}). We included GitHub star counts as an indication of popularity.

\begin{table}[h!]
    \centering
    \resizebox{\linewidth}{!}{%
    \begin{tabular}{|p{1.1cm}||p{1.5cm}|p{1cm}|p{0.8cm}||p{1.5cm}|p{1.5cm}|p{1.5cm}|}
    \hline
    \textbf{Service Mesh} & \textbf{Model} & \textbf{Proxy} & \textbf{GitHub Stars} & \textbf{P99 Latency (seconds)} & \textbf{CPU (cores)} & \textbf{Memory (MiB)\tablefootnote{1 MiB (Mebibyte) is equal to $2^{20}$ bytes}  } \\ \hline \hline
    \textbf{Baseline} & - & - & - & 0.22 & client: 0.08 & client: 152 \\ 
                      &   &   &   &      & server: 0.12 & server: 71 \\ \hline
    \textbf{Istio} & Sidecar & Envoy & \cellcolor{green!30}35.9k & \cellcolor{red!30}$+0.38$ & \cellcolor{red!30}client: $+0.81$ & \cellcolor{red!30}client: $+255$ \\ 
                   &         &       & \cellcolor{green!30}     & \cellcolor{red!30}    & \cellcolor{red!30}server: $+0.87$ & \cellcolor{red!30}server: $+169$ \\ \hline
    \textbf{Istio} & Sidecarless & Ztunnel & \cellcolor{green!30}35.9k & \cellcolor{green!30}$+0.02$ & \cellcolor{orange!30}client: $+0.23$ & \cellcolor{green!30}client: $+26$ \\ 
    \textbf{Ambient} & (node agent) &       & \cellcolor{green!30}     & \cellcolor{green!30}     & \cellcolor{orange!30}server: $+0.23$ & \cellcolor{green!30}server: $+26$ \\ \hline
    \textbf{Linkerd} & Sidecar & Linkerd2 & \cellcolor{red!30}10.6k & \cellcolor{yellow!30}$+0.09$ & \cellcolor{orange!30}client: $+0.29$ & \cellcolor{yellow!30}client: $+62$ \\ 
                     &         & -proxy   & \cellcolor{red!30}       & \cellcolor{yellow!30}    & \cellcolor{orange!30}server: $+0.22$ & \cellcolor{yellow!30}server: $+63$ \\ \hline
    \textbf{Cilium} & Sidecarless & Envoy & \cellcolor{orange!30}20.1k & \cellcolor{orange!30}$+0.22$ & \cellcolor{green!30}client: $+0.12$ & \cellcolor{orange!30}client: $+95$ \\ 
                    & (node agent) &       & \cellcolor{orange!30}      & \cellcolor{orange!30}       & \cellcolor{green!30}server: $+0.08$ & \cellcolor{orange!30}server: $+95$ \\ \hline
    \end{tabular}%
   
    }
    \caption{Comparison of Service Mesh Models Performance 
    (3200 RPS, 1600 connections, intra-node communication). The order of preference is: green $>$ yellow $>$ orange $>$ red. The numbers are relatively ($+$) to the baseline.
    }
    \label{fig:summary-table}
\end{table}

    \begin{table}[h!]
    \centering
    \resizebox{\linewidth}{!}{%
    \begin{tabular}{|p{1.1cm}||p{1.5cm}|p{1cm}|p{0.8cm}||p{1.5cm}|p{1.5cm}|p{1.5cm}|}
    \hline
    \textbf{Service Mesh} & \textbf{Model} & \textbf{Proxy} & \textbf{GitHub Stars} & \textbf{P99 Latency (seconds)} & \textbf{CPU (cores)} & \textbf{Memory (MiB)  } \\ \hline \hline
    \textbf{Baseline} & - & - & - & 0.22 & client: 0.08 & client: 152 \\ 
                      &   &   &   &      & server: 0.12 & server: 71 \\ \hline
    \textbf{Istio} & Sidecar & Envoy & \cellcolor{green!30}35.9k & \cellcolor{red!30}+0.37 & \cellcolor{red!30}client: +0.74 & \cellcolor{red!30}client: +355 \\ 
                   &         &       & \cellcolor{green!30}     & \cellcolor{red!30}    & \cellcolor{red!30}server: +0.74 & \cellcolor{red!30}server: +173 \\ \hline
    \textbf{Istio} & Sidecarless & Ztunnel & \cellcolor{green!30}35.9k & \cellcolor{green!30}+0.03 & \cellcolor{orange!30}client: +0.23 & \cellcolor{orange!30}client: +86 \\ 
    \textbf{Ambient} & (node agent) &       & \cellcolor{green!30}     & \cellcolor{green!30}     & \cellcolor{orange!30}server: +0.24 & \cellcolor{orange!30}server: +97 \\ \hline
    \textbf{Linkerd} & Sidecar & Linkerd2 & \cellcolor{red!30}10.6k & \cellcolor{yellow!30}+0.08 & \cellcolor{orange!30}client: +0.22 & \cellcolor{green!30}client: +53 \\ 
                     &         & -proxy   & \cellcolor{red!30}       & \cellcolor{yellow!30}    & \cellcolor{orange!30}server: +0.15 & \cellcolor{green!30}server: +55 \\ \hline
    \textbf{Cilium} & Sidecarless & Envoy & \cellcolor{orange!30}20.1k & \cellcolor{orange!30}+0.22 & \cellcolor{green!30}client: +0.10 & \cellcolor{orange!30}client: +93 \\ 
                    & (node agent) &       & \cellcolor{orange!30}      & \cellcolor{orange!30}       & \cellcolor{green!30}server: +0.03 & \cellcolor{orange!30}server: +94 \\ \hline
    \end{tabular}%
    }
    \caption{Comparison of Service Mesh Models Performance 
    (3200 RPS, 1600 connections, inter-node communication). The order of preference is: green $>$ orange $>$ red. The numbers are relative (+) to the baseline.}
    \label{fig:summary-table-inter}
    \end{table}

    Our experiments reveal significant performance differences among the service meshes. Enforcing mTLS increased latency across all tested providers, with increases of 166\% for Istio, 8\% for Istio Ambient, 33\% for Linkerd, and 99\% for Cilium. In some tests, Istio's latency increase was almost four times that of Linkerd and more than 6 times that of Istio Ambient. Furthermore, our results highlight the performance benefits of a sidecarless architecture and the use of eBPF. To understand the root cause of Istio's high latency, we discovered that specific steps in request processing, such as HTTP parsing, which were part of the default mTLS implementation, significantly contribute to the overall performance overhead. 

    Our findings can be used by decision-makers to select the most appropriate service mesh provider and architecture for their applications.

   \section{Background}
   \label{sec:background}

\subsection{Kubernetes: Containers, Pods, Nodes and Clusters}
Kubernetes is a widely-used open-source system for automating the deployment, scaling, and management of applications \cite{kubernetes_doc}.
Containers, Pods, Nodes, and Clusters are the core components of Kubernetes. \textit{Containers} are self-contained environments for applications, providing repeatability across diverse cloud and OS infrastructures by decoupling applications from the host infrastructure. A \textit{Pod}, the smallest deployable unit in Kubernetes, can host one or more containers that share storage and network resources. \textit{Nodes}, either physical or virtual machines, provide the infrastructure for running Pods and are managed by the Kubernetes control plane. A cluster consists of a control plane with a set of worker nodes. \textit{Clusters} ensure fault-tolerance and high availability by distributing workloads and managing resources across the system.
Service meshes are built on these Kubernetes components to manage the communication between services within the cluster.

   \subsection{Service Mesh: Architectures}
   \label{sec:Architectures}
\begin{figure}[h]
  \centering
  \begin{subfigure}[b]{0.45\linewidth}
    \centering
    \includegraphics[width=\linewidth]{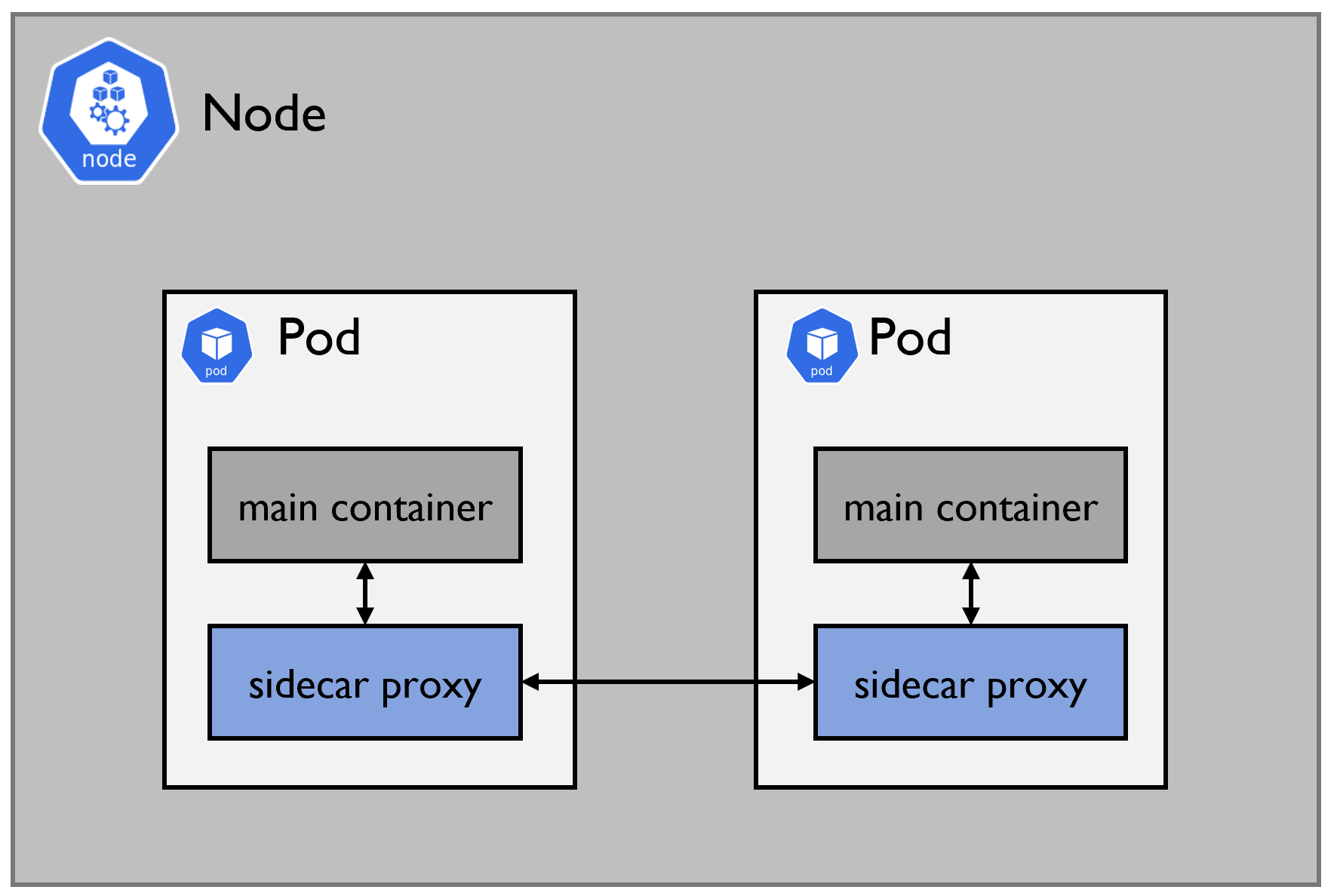}
    \caption{Sidecar Proxy Pattern}
    \label{fig:sidecar1}
  \end{subfigure}
  \hfill
  \begin{subfigure}[b]{0.45\linewidth}
    \centering
    \includegraphics[width=0.85\linewidth]{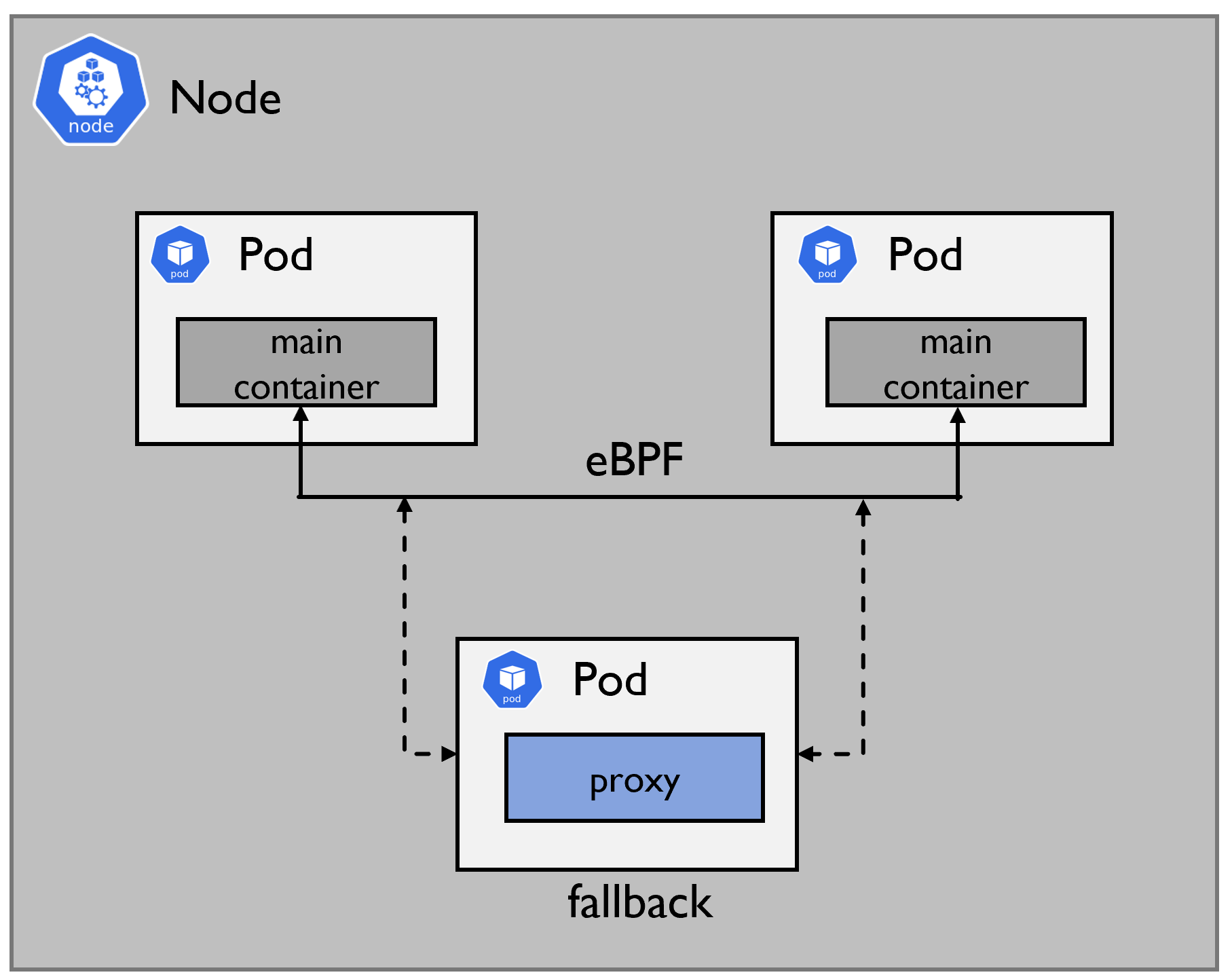}
   \caption{Sidecarless Model}
   \label{Sidecarless Model}
    \label{fig:sidecar2}
  \end{subfigure}
  \caption{Service Mesh Architectural Models}
  \label{fig:sidecar-patterns}
\end{figure}

   There are two main models for managing network traffic in a service mesh: the sidecar pattern and the sidecarless model: The sidecar pattern deploys application components in separate containers for isolation and encapsulation. See Figure \ref{fig:sidecar-patterns}(a). Attached to the main application container, the sidecar shares its lifecycle, starting and stopping with the parent.  
    The sidecar proxy manages all incoming and outgoing network traffic for a service -- handling flow, gathering telemetry, and enforcing policies. The service itself remains unaware of the network, relying on the sidecar. In Kubernetes, the sidecar proxy is an additional container automatically injected into each Pod, managing the main application container's network traffic.

    In the sidecarless model, the proxy container is moved to the host and kernel, eliminating the need for a sidecars for each application pod. This is enabled by eBPF \cite{ebpf}. See Figure \ref{fig:sidecar-patterns}(b). An eBPF-based implementation can use eBPF programs hooked into specific kernel points to redirect packets. A single agent on each Node manages eBPF programs, allowing the Linux kernel to control network access in and out of containers. Features like L3/L4 forwarding, routing, and network policies are handled directly in eBPF, while tasks like connection splicing, rate limiting, or TLS termination fall back to the Node’s central agent for L7 processing, when eBPF is not capable of processing the request.
     
   We chose the tested providers based on their popularity and maturity to ensure the findings of this study are relevant to a wide audience. To better investigate performance factors, we chose service meshes with both similarities and differences. Below, we briefly highlight the key features and differences of the studied service meshes: Istio, Istio-ambient, Linkerd, and Cilium.

    \textbf{Istio} \cite{istio}: A widely adopted open-source service mesh built on Envoy \cite{envoy}.
    It has a data plane with Envoy-based sidecar proxies and a control plane that manages the proxies and traffic routing.
    It was the first Kubernetes-native solution with advanced features like comprehensive analytics.
    
    \textbf{Istio Ambient} \cite{Istio-ambient}: Istio Ambient is a newer, sidecar-less architecture for Istio that aims to reduce the complexity and resources overhead associated with traditional sidecar proxies by using lightweight node-level proxies, utilizing eBPF.
    
   \textbf{Linkerd} \cite{linkerd}: An open-source Kubernetes service mesh under the CNCF, licensed under Apache v2. Architecturally similar to Istio, Linkerd employs a data plane and a control plane, deploying transparent proxies within each Pod to auto-manage traffic. Unlike Istio, Linkerd uses its own lightweight micro-proxy, linkerd2-proxy, instead of Envoy.

    \textbf{Cilium} \cite{cilium}: An open-source project 

    for 
    Kubernetes clusters and other container orchestration platforms. Unlike Istio and Linkerd, Cilium adopts a unique service mesh architecture by allowing the mesh to operate entirely without sidecars. Instead of deploying proxy containers with each pod, it relocates the functionality to the host and kernel, utilizing eBPF.

  \subsection{mTLS: Network Security}
Transport Layer Security (TLS) is a widely used security protocol that ensures privacy and data integrity. Proposed by the Internet Engineering Task Force (IETF) in 1999, the latest version, TLS 1.3, was released in 2018 \cite{tls2018}. 
It is typically layered atop the Transmission Control Protocol (TCP) to secure application protocols like HTTP but can also work with UDP. TLS uses asymmetric cryptography to verify the server’s identity and symmetric key cryptography to encrypt the data exchanged between client and server, preventing man-in-the-middle attacks and keeping intercepted messages unreadable.

Mutual Transport Layer Security (mTLS) is very similar to the TLS protocol. While in TLS, the client verifies the server’s identity by asking for and validating its certificate. In mTLS there is an additional step involved in which the server also asks for the client's certificate and verifies it at their end. A secure connection for data transfer is only established when both client and server successfully authenticate themselves and verify each other’s certificates. 
As companies shift to zero-trust security\footnote{The zero-trust security model is based on guidelines established by the National Institute of Standards \& Technology (NIST). The NIST 800-207 \cite{nist-zero-trust} publication defines the standard for zero-trust practices. In this model, all users, whether internal or external, must undergo continuous authentication and verification before accessing data.}, mTLS is becoming very popular, providing secure communication policies between microservices.

\section{Related Work}
Performance evaluations of service mesh have been conducted by the providers themselves: Istio \cite{istio_performance}, Linkerd \cite{linkerd_benchmarks}, Cilium \cite{cilium_performance}, and \cite{istio_vs_linkerd_1,istio_vs_linkerd_2,istio_vs_linkerd_3}. 
Repeating these tests from a third-party perspective is crucial to validate their results and offer deeper insights.
Several studies have explored the performance overhead of service meshes. 
\cite{zhu2022dissecting} developed MeshInsight, profiling components and predicting overhead based on configuration and workload. \cite{saleh2022empirical} examined Istio's traffic management features (e.g., circuit breakers and retries) impact on performance through experiments with varying configurations.
\cite{challenges_perf_sm_edge} found significant performance impacts of service meshes in edge environments. 
Recent benchmarks \cite{cilium_service_mesh} evaluated a sidecar-free model (Cilium), showing significant latency improvements compared to other service meshs by avoiding the overhead of running two proxies between connections.
\cite{duque2022qualitative} focused on performance-demanding workloads and concluded that current service mesh architectures are inadequate, calling for research into more efficient designs and improved quality of service.

The performance impact of TLS has been studied extensively. In 1999 \cite{tls-Apostolopoulos} found that the overhead due to TLS can decrease the number of HTTP transactions handled by up to two orders of magnitude. 
In 2018, \cite{MACHNIK2018} evaluated the additional overhead of mutual authentication instead of just server-side authentication and showed that its further effect is minimal both in latency and throughput, either with or without session reuse. \cite{MACHNIK2018} stated that 
TLS may only incur 12-40\% penalty degradation thanks to session reuse.
\cite{perf_analysis_zero_trust} found Istio's Ingress Gateway more performance efficient than Kubernetes' Load Balancer in a Zero-Trust multi-cloud environment.

\section{Methodology}

Our methodology involves comparing native Kubernetes (referred to as the baseline), Kubernetes with mTLS (referred to as the baseline with mTLS), and a specific service mesh technology with mTLS.  Our experiments are limited to a single cluster, but we compare inter-node (communication between different nodes in the cluster) and intra-node (communication between pods  within the same node) scenarios.

\ignoreA{ Creating a realistic and representative testing environment is crucial for obtaining results that accurately reflect real-world scenarios. Our testing infrastructure was designed to mimic a production-like microservices ecosystem while allowing for controlled and repeatable experiments.
In this section, we describe the methodology employed in setting up the testing infrastructure and designing the tests. }

\label{chap:perf-benchmarking}
We designed a representative microservices testing environment on an OpenShift Kubernetes Cluster hosted on GCP VMs (see specification at Table \ref{tab:machine-specs} ). 
The cluster consisted of three node groups: \textit{Control plane} (Kubernetes system pods), \textit{System} (OpenShift components, monitoring tools, etc), and \textit{Workload} (load generator, server). Separating these nodes isolated the testing subjects and minimized interference.

    \begin{table}[h!]
   \centering
   \begin{tabular}{|l|l|}
   \hline
   \textbf{Machine type} & n2d-highcpu-8 \\ \hline
   \textbf{Kubernetes Version} & 1.24.6 \\ \hline
   \textbf{Node CPU} & 8 vCPU \\ \hline
   \textbf{Node Memory} & 8 GB \\ \hline
   \textbf{OS} & Fedora CoreOS 36 \\ \hline
   \textbf{Kernel Version} & 6.08-200.fc36.x86\_64 \\ \hline
   \textbf{Container Runtime} & cri-o://1.24.3 \\ \hline
   \end{tabular}
   \caption{Node Specifications}
   \label{tab:machine-specs}
   \end{table}

For the workload generation, we utilized Fortio~\cite{fortio} to generate simulated network traffic. Fortio operates at a specified RPS (Requests Per Second), records execution time histograms, and calculates percentiles. \ignoreA{ It can run for a set duration, a fixed number of calls, or continuously until interrupted, maintaining a constant target RPS or maximum load per connection/thread.}
For simulating a server, we implemented a simple HTTP server in Go using the \texttt{net/http} library~\cite{go-http-lib}. \ignoreA{To enforce mTLS in native Kubernetes without service mesh, we utilize the \texttt{crypto/tls} library~\cite{go-tls-lib}. }To simulate server processing time, we added a 200ms delay before returning responses.

To evaluate performance overhead, we focused on \textit{Latency}, \textit{CPU Usage}, \textit{Memory Usage}, and \textit{Error Rate}. Fortio reports latency metrics exclusively in the form of percentiles. Therefore, we analyzed the 99th and 50th percentile latencies in this study. As the trend from the results from both percentiles were nearly identical, we chose to focus solely on the 99th percentile in this paper. We show in the Figures only the server side performance however the client side exhibits a similar trend.  For the complete set of results (with client-side measurement, and 50th percentile) see \ref{data} and for the exact configuration flags see \ref{sec:config-of-service-mesh}. 

For service meshes that have the sidecar pattern approach, we enabled the automatic mechanism for injecting the sidecar proxy. In addition, We configured each service mesh to enable or disable mutual TLS (mTLS) as required. In Cilium, the configuration  enables IPSec or WireGuard for the inter-node communication and there is no authentication in the intra-node communication. For identity management, we deployed the SPIRE server. 

\subsection{Test Scenarios}\label{sec:test-scenarios}
We assessed each service mesh's performance overhead by conducting tests starting with a baseline without any service mesh, then repeating the tests under identical conditions with each service mesh installed to quantify the overhead. Tests were executed at constant request rates, for 5 minutes each, and repeated multiple times to ensure result stability, indicated by low variance. 
Source code is available on GitHub \cite{git}.

We designed several test scenarios to evaluate the performance impact of different service mesh configurations.

\subsubsection{Baseline scenario} The baseline test measured the system's performance without any service mesh, serving as a reference point for subsequent comparisons. Traffic routing relied on Kubernetes' native networking with \texttt{kube-proxy}\cite{kube_proxy}. 

\subsubsection{Baseline mTLS scenario} This test is similar to the baseline test but with mTLS enforced between the client and server using Go's \texttt{crypto/tls} library~\cite{go-tls-lib}. Certificates were generated using OpenSSL~\cite{openssl}. 

\subsubsection{Sidecar Pattern scenario} In this scenario, relevant to Istio and Linkerd, we injected sidecar proxies into the pods., and configured them to enforce mTLS communication. \ignoreA{ The test topology is as described in sec~\ref{sec:Architectures} but with mTLS.}

\subsubsection{Sidecarless Pattern scenario} For Cilium and Istio Ambient, we evaluated the sidecarless pattern, where authentication and encryption are handled by a per-node agent without sidecar proxies. The proxies were configured to enforce mTLS communication.

\newcommand{\ofek}[1]{\textcolor{blue}{\textbf{[Ofek: #1]}}}

\section{  Experimental Results}
   \label{chap:tests-results}

\ignoreA{   This section presents the performance evaluation results. The metrics extracted from each test scenario, as described in Section \ref{sec:performance-metrics}, were used to assess the system under various conditions. Each chart consists of three clusters of bars, with each cluster representing a different test case under varying load levels, as detailed in Section \ref{sec:performance-metrics}.}

   
   \subsection{ Baseline Test: mTLS impact }
   \label{sec:baseline-test-results}
    Prior to analyzing the impact of Service Mesh implementations, we first established a baseline to understand the effect of mTLS on performance in the absence of a service mesh. We conducted 5-minute tests with 160, 1600, and 6400 concurrent connections at 320, 3200, and 12,800 RPS, respectively (2 RPS for each connection).
   
\begin{figure}[h]
   \includegraphics[width=\linewidth]{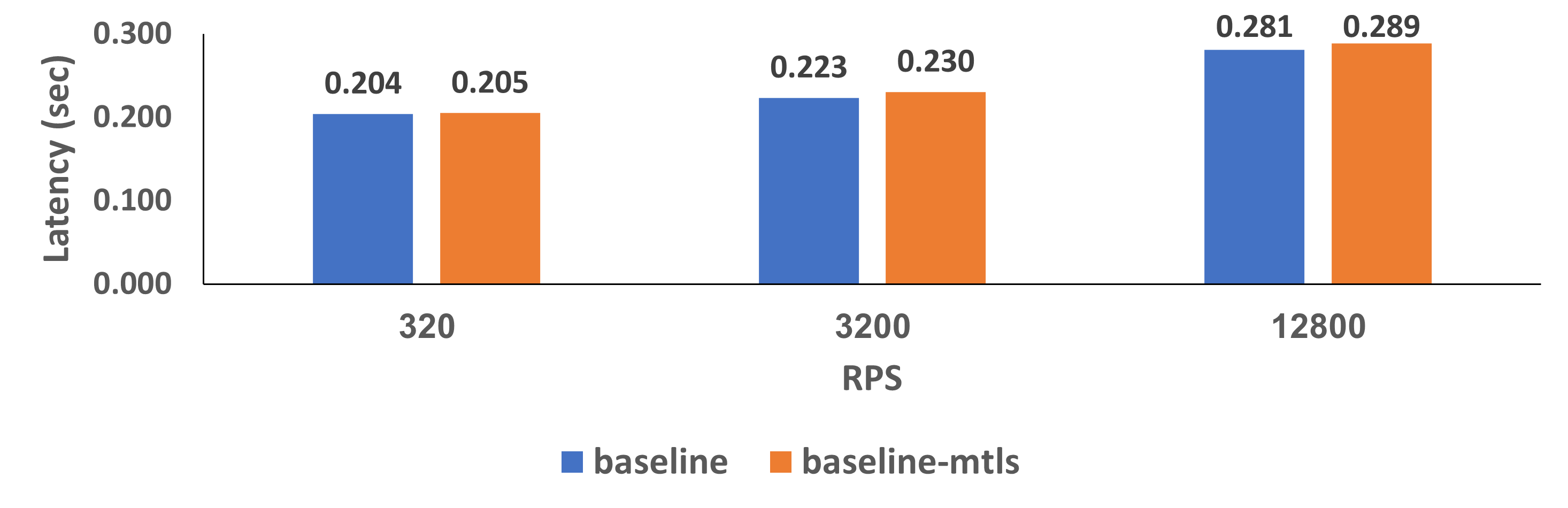}
   \centering
   \caption{Baseline tests P99 Latency}
   \label{fig:baseline-tests-latency}
   \end{figure}

  
   Figure \ref{fig:baseline-tests-latency} illustrates the latency  from the baseline test compared to the baseline-mTLS test.
   \\
   The latency increase in the baseline-mTLS test is relatively low. There is up to 3\% increase in the 99th percentile. Even under high load conditions (larger number of connections), the latency overhead introduced by mTLS was relatively small. The results for the 50th percentile were similar.

\begin{figure}[h]
   \centering
   \begin{subfigure}[b]{\linewidth}
      \centering
      \includegraphics[width=0.97\linewidth]{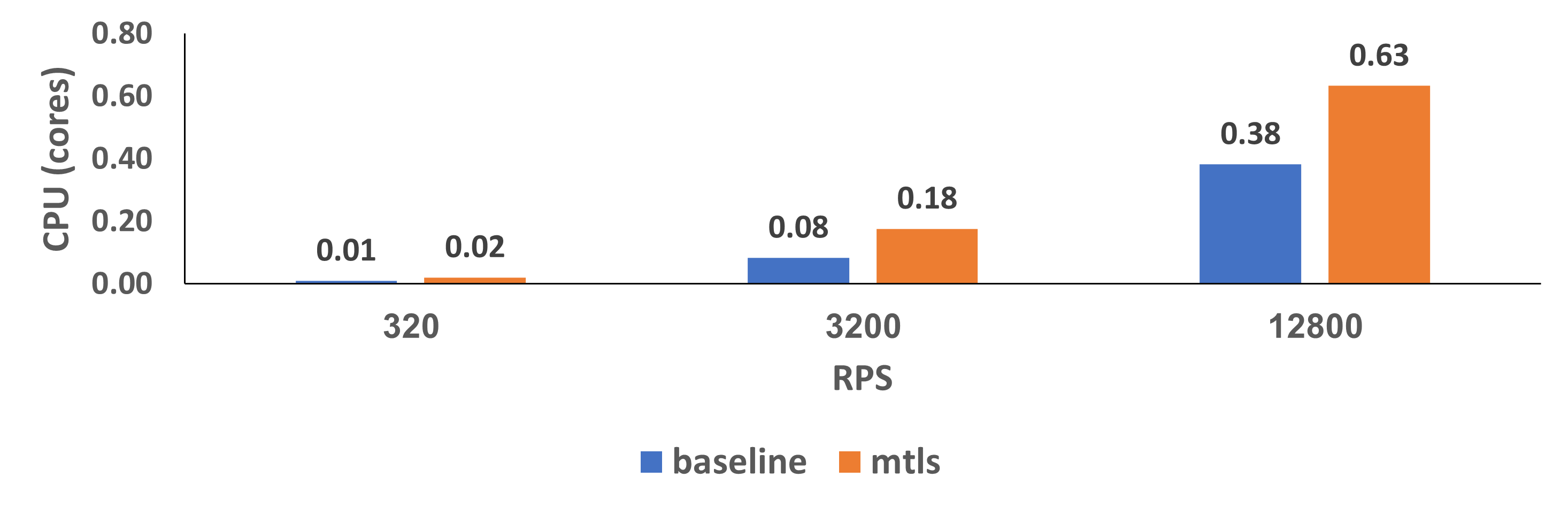}
      \caption{CPU Usage}
      \label{fig:baseline-tests-resources-cpu}
   \end{subfigure}

   \begin{subfigure}[b]{\linewidth}
      \centering
      \includegraphics[width=0.97\linewidth]{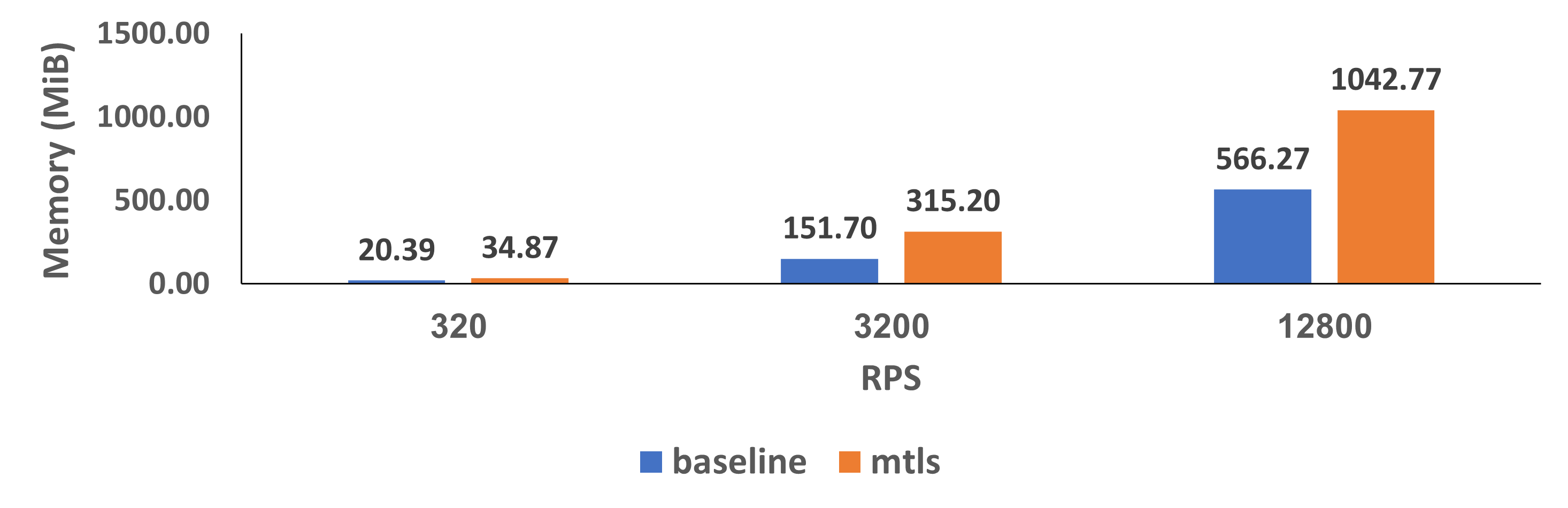}
      \caption{Memory Usage}
      \label{fig:baseline-tests-resources-memory}
   \end{subfigure}
   
   \caption{Baseline Resource Tests - Client}
   \label{fig:baseline-tests-resources}
\end{figure}

   However, Fig \ref{fig:baseline-tests-resources}, shows that enabling mTLS led to a notable increase in resource usage, with CPU and memory consumption increasing by up to 100\% in client pods (and similarly in server pods - see \ref{data}).

   \subsection{Service Mesh : mTLS impact }
   \label{sec:sm-load-results}
    In each test, we installed the selected service mesh in the cluster and 
   enforced using mTLS by the service mesh and conducted 5-minute tests with 160, 1600, and 6400 concurrent connections at 320, 3200, and 12,800 RPS, respectively (2 RPS for each connection).

    \begin{figure}[h]
       \centering
       \includegraphics[width=\linewidth]{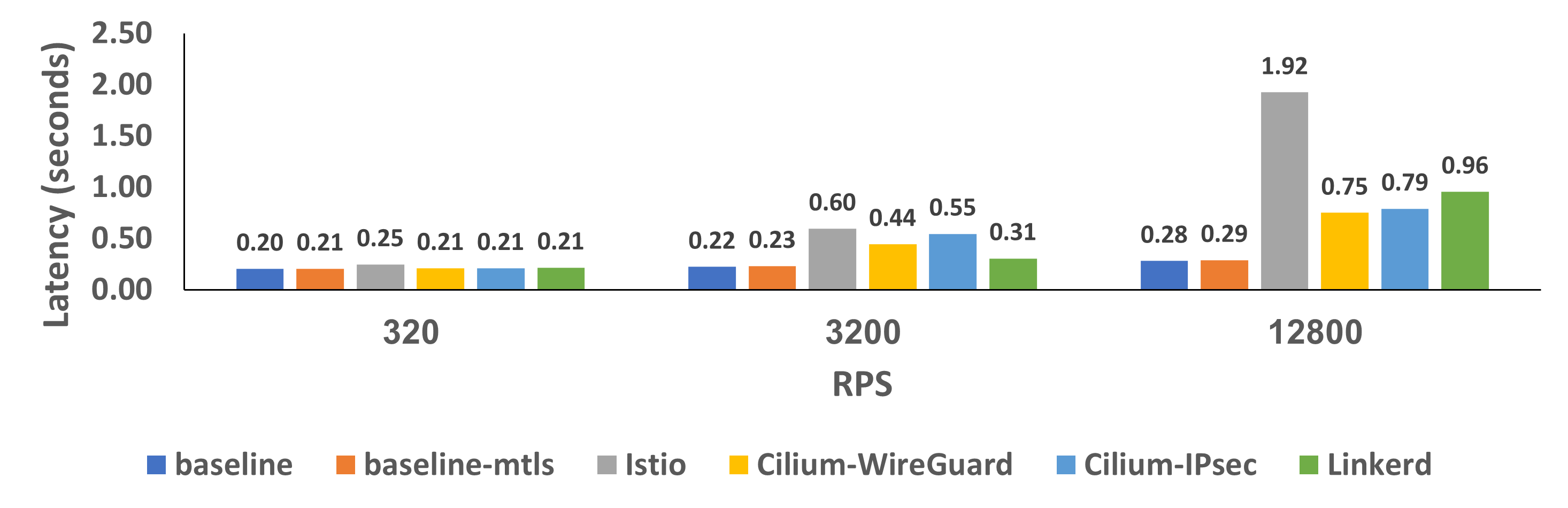}
       \caption{P99 Latency as function of load}
       \label{fig:sm-tests-latency}
   \end{figure}

   \ignoreA{\begin{figure*}[h]
   \centering
   \begin{minipage}{0.49\linewidth}
       \centering
       \includegraphics[width=\linewidth]{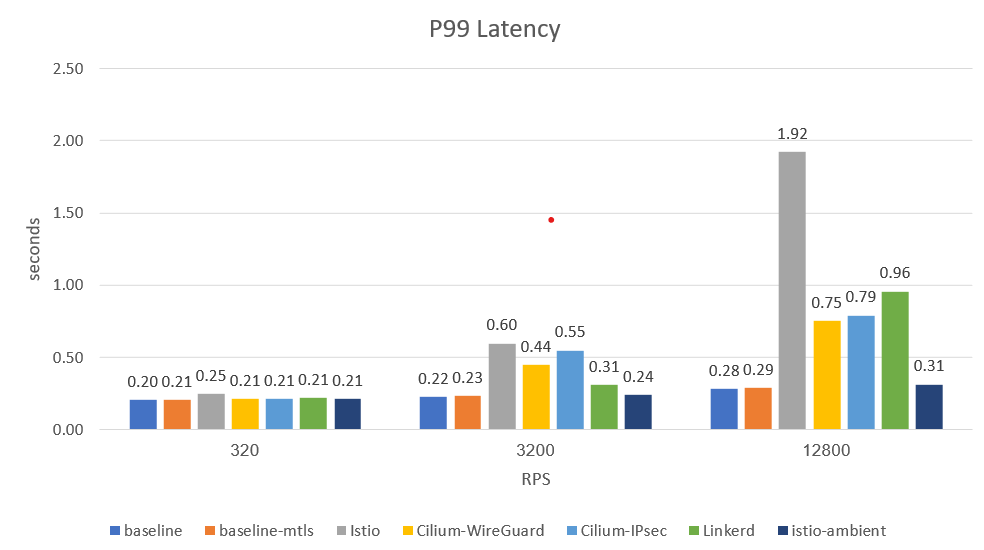}
       \subcaption{99th Percentile Latency}
   \end{minipage}
   \hfill
   \begin{minipage}{0.49\linewidth}
       \centering
       \includegraphics[width=\linewidth]{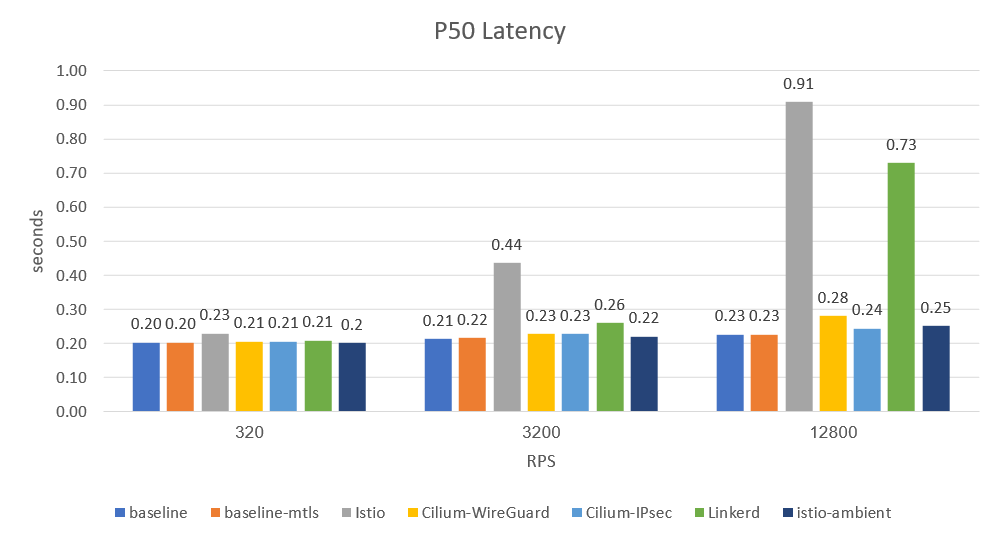}
       \subcaption{50th Percentile Latency}
   \end{minipage}
   \caption{Service Mesh tests latency}
   \label{fig:sm-tests-latency}
\end{figure*}}

    As can be seen in Fig \ref{fig:sm-tests-latency}, a significant latency increase was measured in all service meshes with high variance between them. Istio had the highest latency increase, and Istio Ambient had the lowest latency. Moreover, as the number of RPS increases, the difference between the service meshes grows as well. In the test with the highest load (12,800 RPS), the load generator could not even achieve the target RPS with the Istio sidecar due to the high latency (achieved 6868 RPS).

\begin{figure}[h]
   \centering
   \begin{subfigure}[b]{\linewidth}
      \centering
      \includegraphics[width=0.95\linewidth]{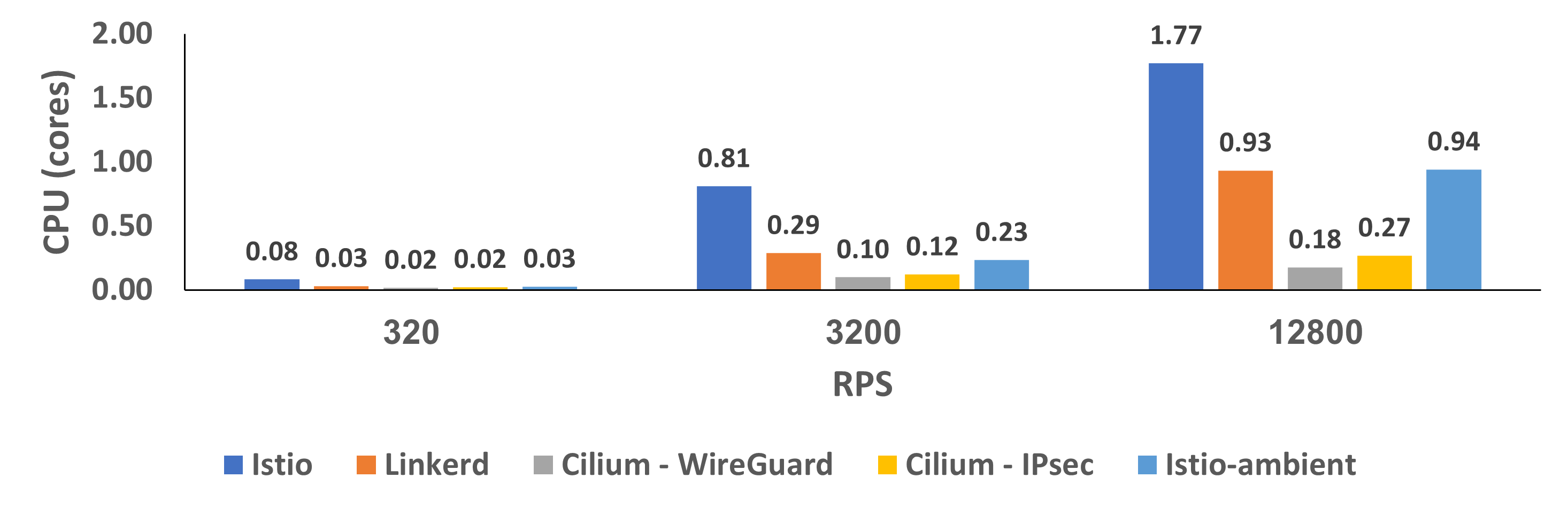}
      \caption{CPU usage}
      \label{fig:sm-tests-cpu}
   \end{subfigure}

   \begin{subfigure}[b]{\linewidth}
      \centering
      \includegraphics[width=0.95\linewidth]{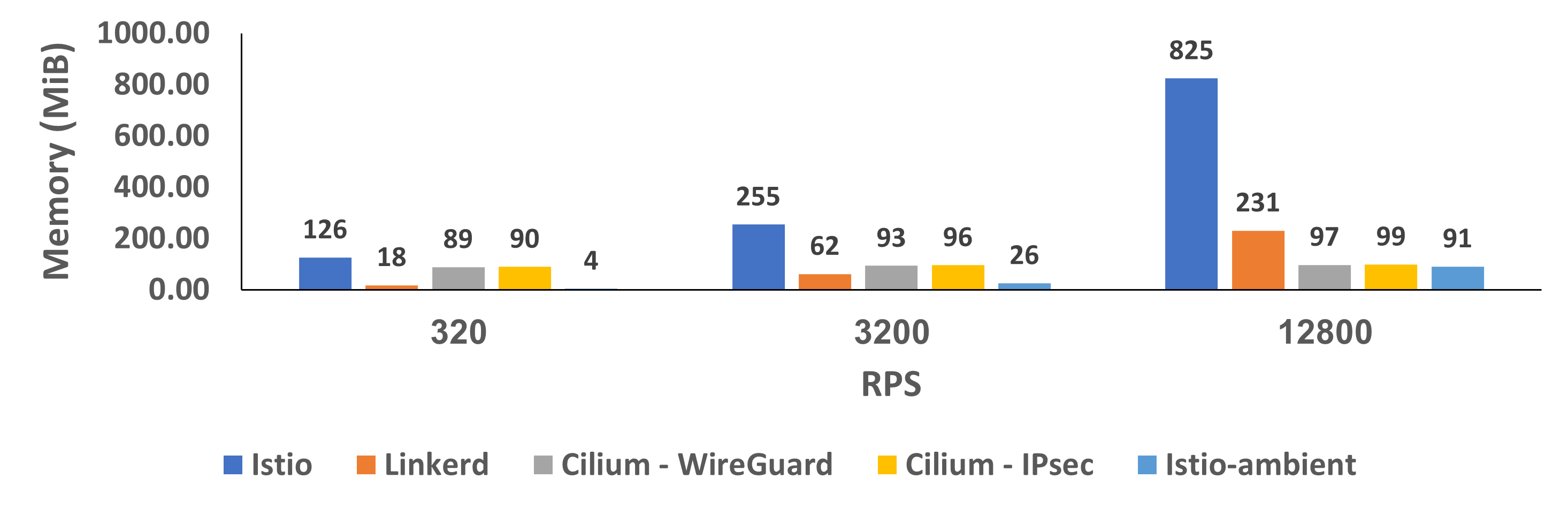}
      \caption{Memory usage}
      \label{fig:sm-tests-memory}
   \end{subfigure}

   \caption{Service Mesh's proxies CPU and Memory Usage - Client}
   \label{fig:sm-tests-resource}
\end{figure}

    \footnotetext{It represents sidecar proxies' resources in Istio and Linkerd, and the central node agent's resources in Cilium and Istio Ambient.}
    Fig \ref{fig:sm-tests-resource} shows that Cilium and Istio-Ambient had the best performance in terms of CPU consumption and memory consumption respectively, while Istio was the worst in terms of CPU and memory consumption.

\begin{figure}[h]
   \includegraphics[width=\linewidth]{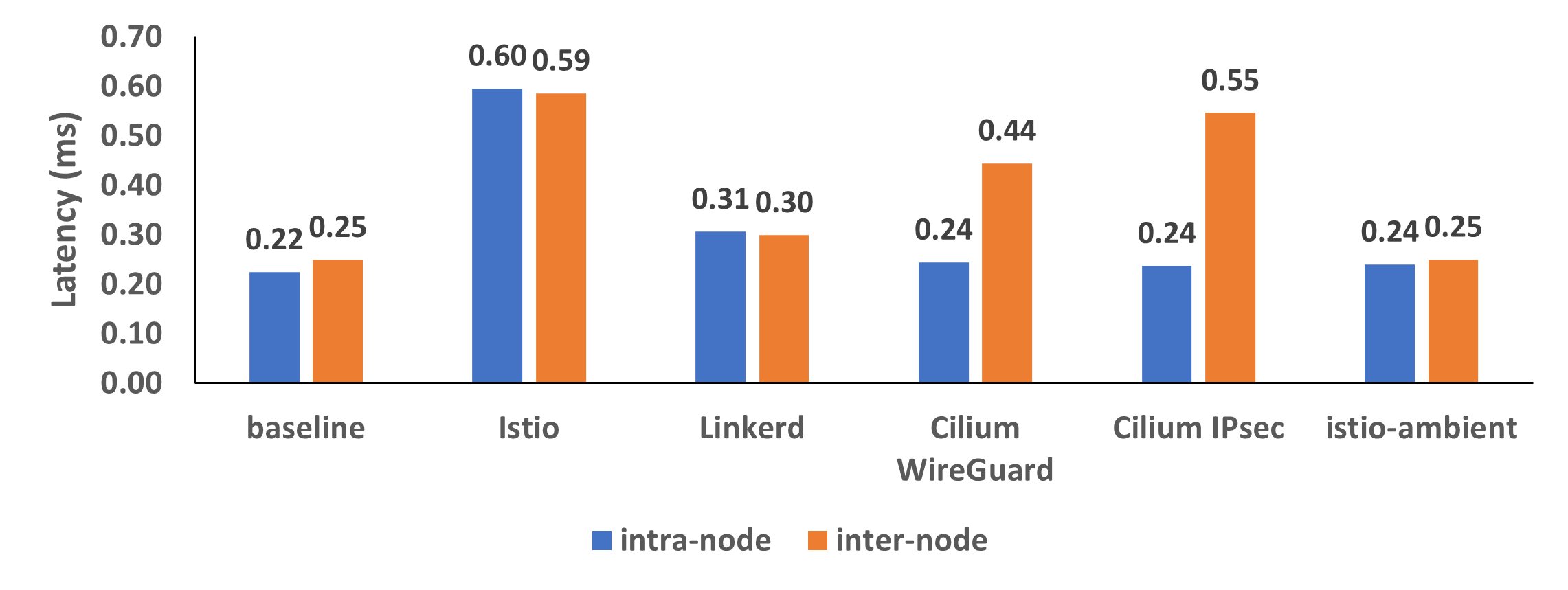}
   \centering
   \caption{Inter-node vs Intra-node communication P99 latency}
   \label{fig:inter-node-tests-latency}
   \end{figure}

   Fig. \ref{fig:inter-node-tests-latency} shows the latency difference between inter-node communication (communication between different nodes) and intra-node communication (communication within the same node). It demonstrates a notable latency difference for Cilium, while Istio and Linkerd yielded similar results. While mTLS was enabled in all cases, Cilium, unlike Istio and Linkerd, disables network encryption for intra-node traffic by design. This behavior, consistent across both WireGuard and IPsec configurations, leads to security implications but with a significant performance improvement.

   \subsection{Impact of Configuration on Istio Throughput}
   \label{sec:sm-throughput-test-results}
   In this test, we focus on Istio, which shows the highest latency and memory consumption with the number of connections compared to the baseline. Our goal is to better understand the impact of different configurations on the observed performance. 
    We tested throughput on several scenarios, where in each scenario, the load generator was configured to achieve maximum RPS, initiating new requests immediately after previous ones were completed. We check the impact of enabling the combination of three independent features in Istio: (1) mTLS, (2)Metrics - telemetry collection and (3) Http parsing. 
    
    The scenarios we tested are:
\begin{inparaenum}[(i)]
    \item \textit{mTLS} -  mTLS is enabled with Metrics and Http parsing.
    \item \textit{Plaintext} - mTLS is disabled but Metrics and Http parsing are enabled.
    \item \textit{no-Metrics} - mTLS, Metrics are disabled and Http parsing is enbaled,
    \item \textit{TCP} - mTLS and HTTP Parsing are disabled, but Metrics is enabled.
    \item {TCP-no-Metrics} - all features are diabled: mTLS and HTTP Parsing and Metrics are disabled.
\end{inparaenum}

\begin{figure}[h]   \includegraphics[width=\linewidth]{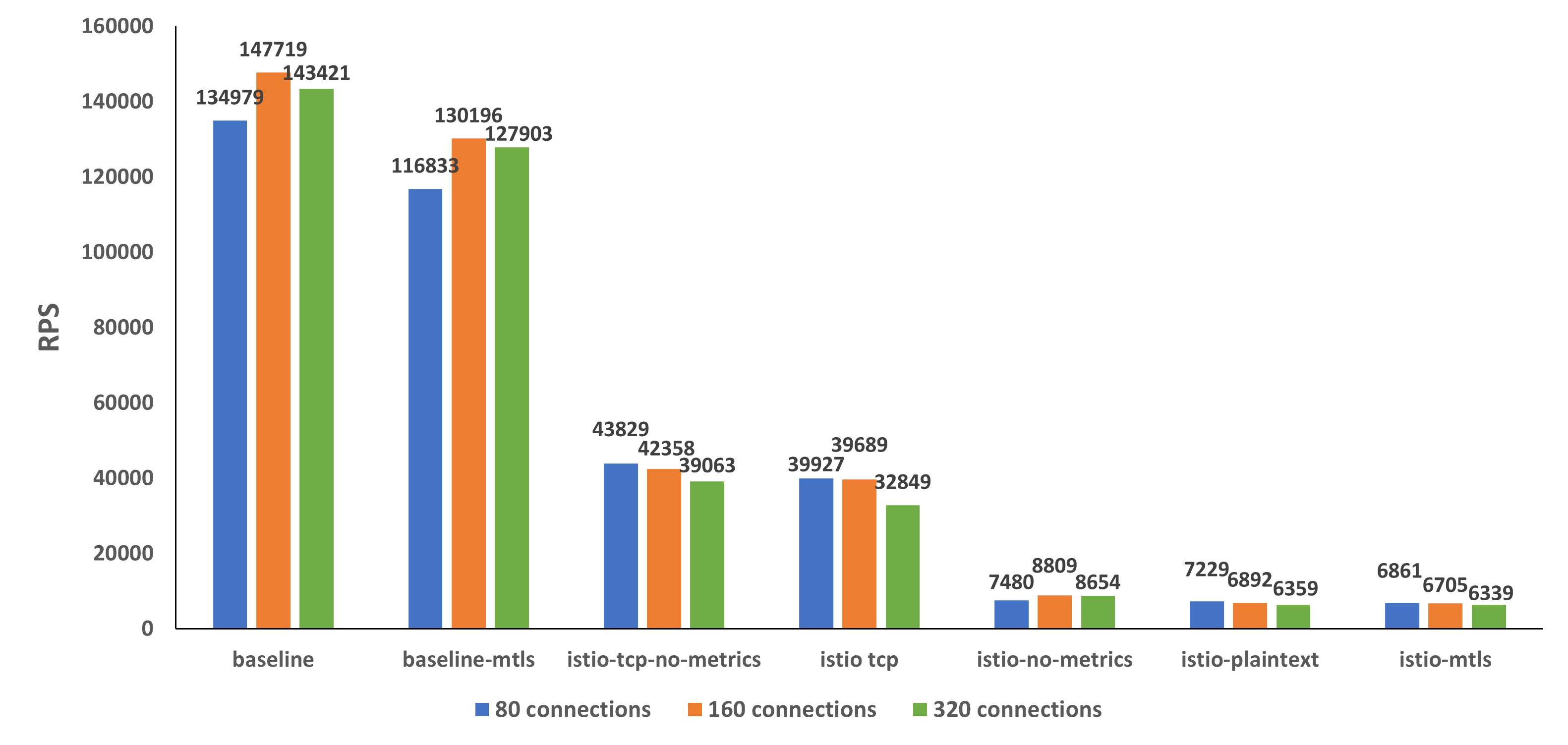}
   \centering
   \caption{Istio throughput with different configurations comparing to the baseline}
   \label{fig:sm-throughput}
   \end{figure}

    As shown in Fig \ref{fig:sm-throughput} adding Istio’s sidecar with mTLS caused a 95\% decrease in throughput compared to the baseline (Kubernetes without Istio). Disabling mTLS produced similar results, indicating that the high latency is not due to the mTLS enforcement but due to the sidecar and extra functionalities. Disabling metrics collection only slightly improved throughput. Istio-tcp, where HTTP parsing was also disabled, increased throughput nearly fivefold compared to Istio with only mTLS disabled, indicating the high resource consumption of HTTP parsing compared to mTLS and metrics collection.

    \ignoreA{The results indicate that disabling each feature increases throughput. Although we tested only a few functionalities, it is reasonable to say that any addition to the sidecar's processing path can potentially affect latency and throughput. While some features introduce more latency than others, their combined effect significantly increases latency, especially under high loads. According to Istio's documentation, collecting telemetry data does not add to a r
    equest's completion time. However, since the worker is occupied with the current request, it cannot immediately handle the next one, leading to increased queue wait times and affecting average and tail latencies.}

\subsection{Memory Consumption Analysis}
   
    \begin{figure}[h]
   \includegraphics[width=\linewidth]{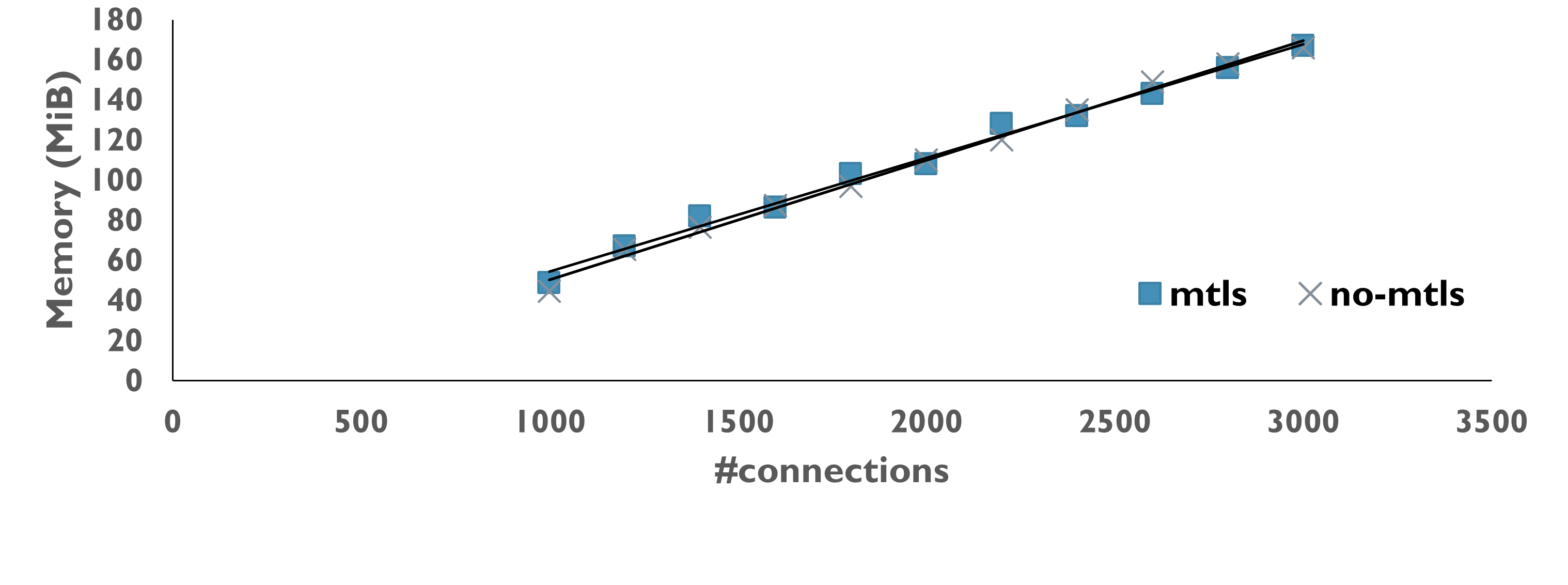}
   \centering
   \caption{Istio-ambient: Memory Consumption by Number of Connections. Measuring client-side with  inter-node communication.}
   \label{fig:mem_cons}
   \end{figure}

To further investigate the impact of memory consumption differences between service mesh implementations, we tested whether memory usage scales with the number of connections and if it is impacted by the mTLS. We chose to focus on Istio Ambient this time, since it gave the best memory results. In this experiment, we maintained a constant request rate of 3200 RPS while increasing concurrent connections from 1000 to 3200. As shown in Fig. \ref{fig:mem_cons}, memory consumption increased linearly with the number of connections, indicating each connection adds a constant memory overhead. The coefficient of determination 
($R^2$) 
was 0.99, confirming a strong linear relationship. The results with the server and with intra-node communication were similar.

However, enabling mTLS did not show an increase in memory consumption  and the two graphs of memory consumption with MTLS and without are similar as connections grew. This indicates that while connection count affects total memory usage, the overhead from mTLS remains relatively constant regardless of the number of connections.

    \subsection{Discussions }

\subsubsection{Istio}
Istio exhibited the highest latency increase, with a 166\% rise at 3,200 RPS, and failed to reach the target at 12,800 RPS due to its sidecar model. This architecture, which adds a proxy for each pod and extra network hops, also resulted in the highest CPU and memory overhead.

       \subsubsection{Istio Ambient}
    
    Istio Ambient showed the best latency performance, with only an 8\% increase at 3,200 RPS and low latency even at 12,800 RPS, outperforming all the others. It was also the most memory-efficient, eliminating per-pod proxies. 
    In terms of CPU usage, Istio Ambient improved significantly over traditional Istio and was similar to Linkerd, though less efficient than Cilium. 
    Overall, Istio Ambient offers a scalable solution with relatively low resource usage while maintaining a robust feature set.

    \subsubsection{Linkerd}
    In terms of latency, Linkerd performed moderately. 
    It lagged behind Istio Ambient across all load levels, and behind Cilium at 12,800 RPS, indicating that while Linkerd is efficient, its sidecar-based architecture introduces more overhead than Istio Ambient’s node-level model.
    For CPU consumption, Linkerd performed similarly to Istio Ambient, better than Istio but not as efficient as Cilium. In terms of memory consumption, Linkerd ranked second best, just behind Istio Ambient. However, in large-scale environments with many pods, Linkerd’s sidecar model may increase resource consumption due to the need for proxy for each pod.

    \subsubsection{Cilium}
    In terms of latency, Cilium showed moderate performance at  3,200 RPS with 99\% latency increase, while
    it outperformed Linkerd at 12,800 RPS, showing better efficiency with higher traffic.
    Cilium performed best in terms of CPU consumption, with its sidecarless model eliminating the need for a proxy per pod, making it highly scalable for large clusters. 
    Its memory consumption was higher than both Linkerd and Istio Ambient. Yet, in clusters with more pods per node, Cilium’s memory efficiency could surpass Linkerd’s, which might make it more favorable in such environments, this needs to be checked in future work. Note, however, that by design, Cilium does not encrypt intra-node traffic, which may be a necessary feature in some cases, but this approach leads to significantly lower latency, reaching latency similar to Istio-Ambient.

\subsubsection{ mTLS protocol Impact on network performance}

   The baseline results show that mTLS has minimal impact on latency in Kubernetes service-to-service communication, but significantly increases CPU and memory usage. This suggests that resource overhead varies by implementation and programming language, and the increased resource consumption should not be overlooked. Additionally, the low latency impact may not apply universally across different scenarios or network topologies, and in the future work, cases where the client and server are in different clusters should be tested. \ignoreA{ The baseline test results indicate that mTLS adds minimal latency overhead but does impact resource usage, which might vary across different libraries and programming languages. However, this does not necessarily apply to all scenarios.  We tested service-to-service communication within the same cluster region. Future studies should explore the potential latency increase from mTLS handshakes in remote regions.}

    Our experiments show that shifting mTLS implementation to the service mesh significantly affects latency and resource consumption. The throughput test in the Istio example reveals this impact stems not from inefficient mTLS but from adding a new network entity, with more functionalities, each contributing to performance overhead.

   \section{Conclusions}
  
  \ignore{ \vspace{-1em}
   \begin{figure}[h]
   \includegraphics[width=\linewidth]{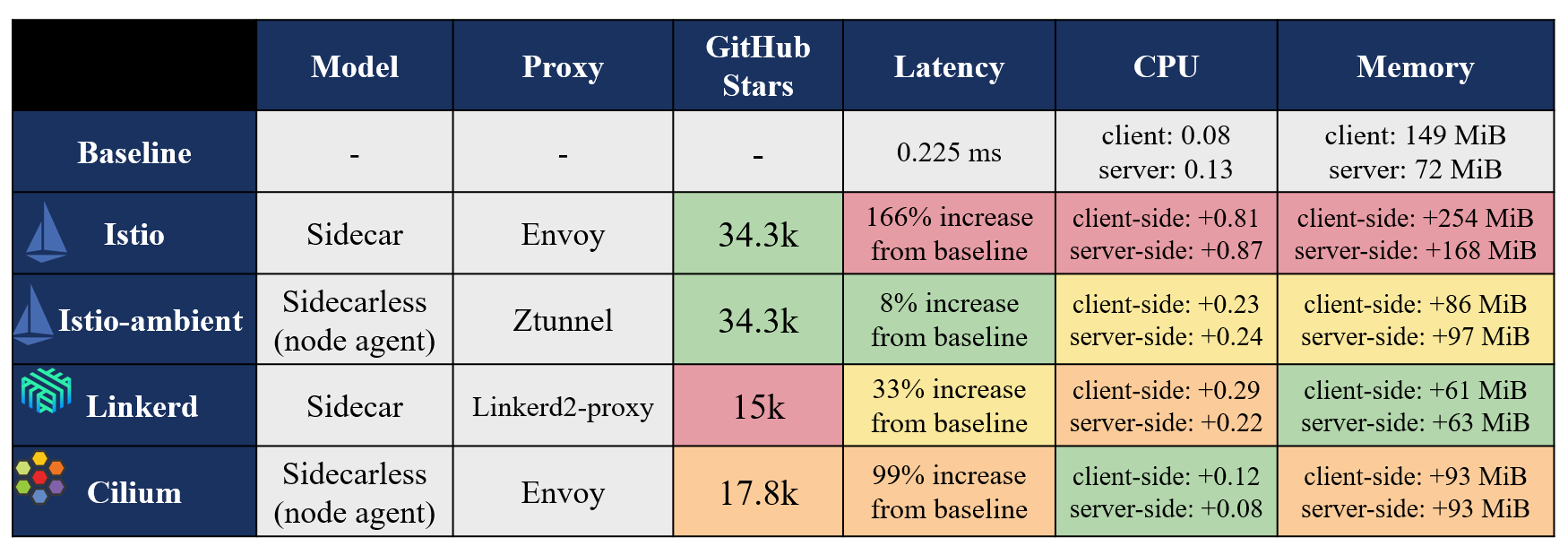}
   \centering
   \caption{Summary of Experimental Results (3200 RPS test).}
   \label{fig:summary-table}
   \end{figure}}

  The goal of this paper was to determine the best-performing service mesh in the mTLS case. However, the results reveal the complexity of evaluating performance and trade-offs, partly due to many hidden features and different functionality in the default settings of the different service mesh frameworks. Nonetheless, it is evident that a sidecarless architecture offers a performance advantage.
   
   \ignoreA{ The performance overhead of service meshes is critical, especially under high loads.  Our analysis reveals significant variations due to architectural and implementation differences, requiring trade-offs. Istio provides rich features but incurs high latency and resource overhead from its sidecar model. Linkerd offers simplicity with lower overhead, though still affected by sidecar costs. Cilium's sidecarless design excels in CPU efficiency but uses more memory, while Istio Ambient balances low latency and memory efficiency, making it suitable for large-scale clusters, though slightly less CPU efficient than Cilium.
    There is no correlation between popularity (GitHub stars, Table \ref{fig:summary-table}) and performance, suggesting that developers prioritize features and community support over performance. 
    This study highlights the importance of aligning service mesh selection with the specific requirements of the application.
    and performance goals, while giving tools to decision-makers for applying the correct service mesh for their needs.}

\section*{Acknowledgment}	
This research was partly supported by RedHat Research.

\bibliographystyle{IEEEtran}
\bibliography{refs}


\appendix
\subsection{P99 Latency and Server's Resource Consumption}
\label{data}
\begin{figure}[h]
   \includegraphics[width=\linewidth]{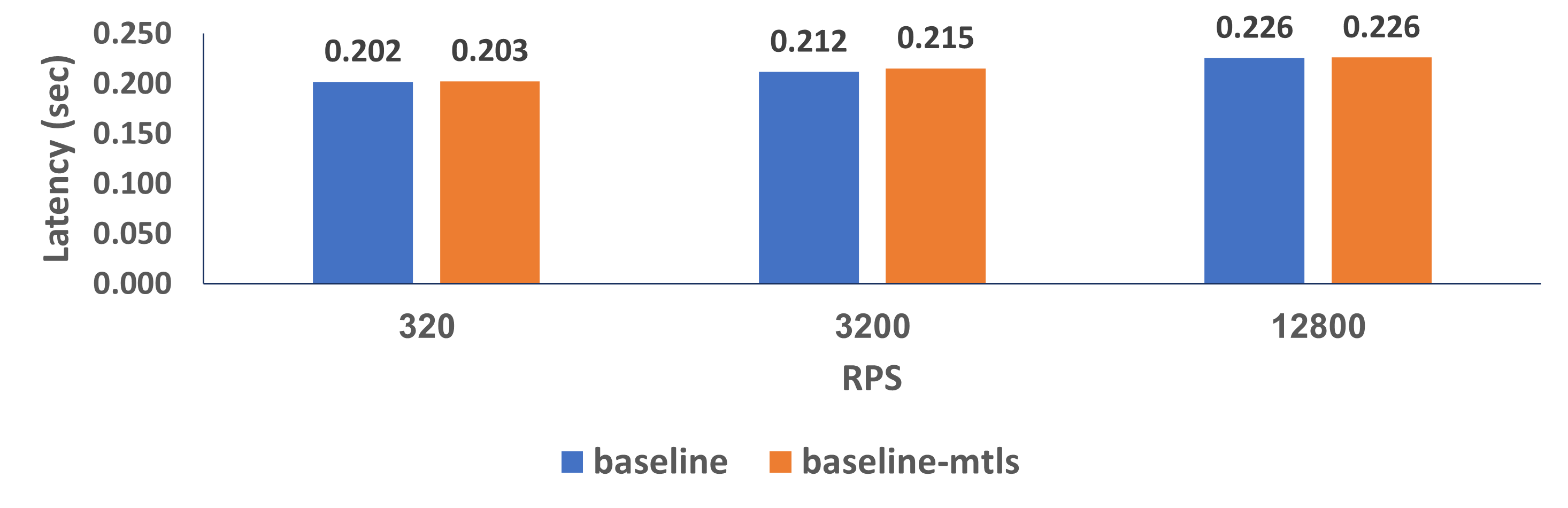}
   \centering
   \caption{Baseline tests P50 Latency}
   \label{fig:baseline-tests-latency-p50}
   \end{figure}
\begin{figure}[h]
   \centering
   \begin{subfigure}[b]{\linewidth}
      \centering
      \includegraphics[width=0.97\linewidth]{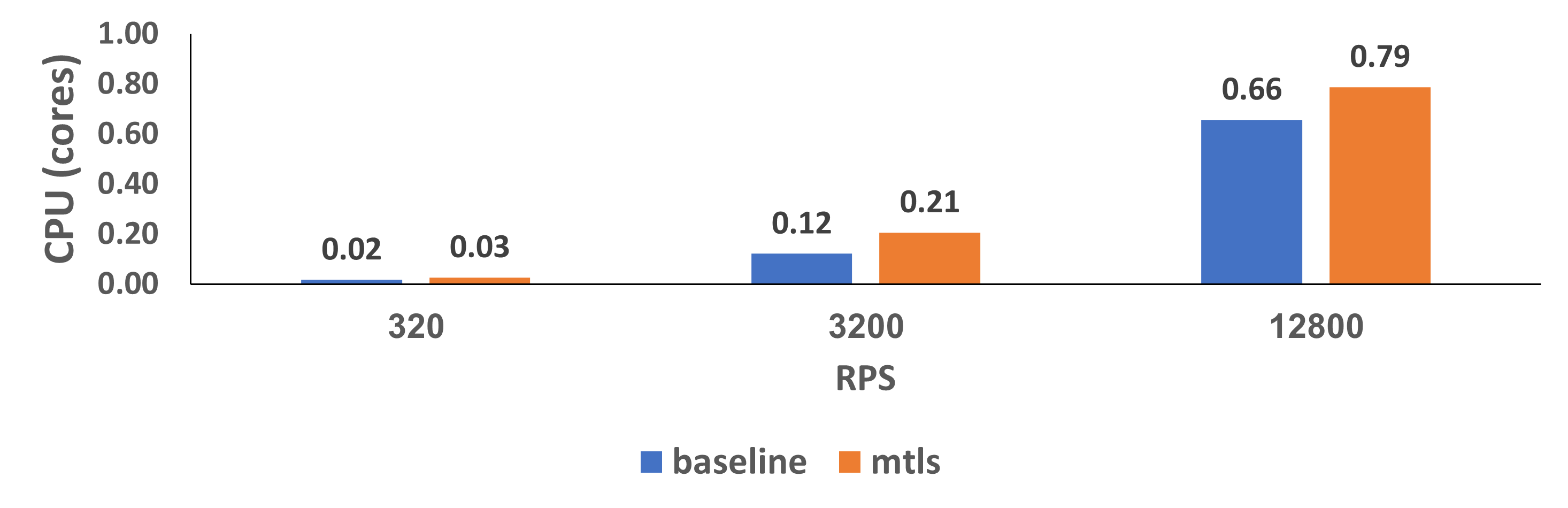}
      \caption{CPU Usage}
      \label{fig:baseline-tests-resources-cpu-server}
   \end{subfigure}

   \begin{subfigure}[b]{\linewidth}
      \centering
      \includegraphics[width=0.97\linewidth]{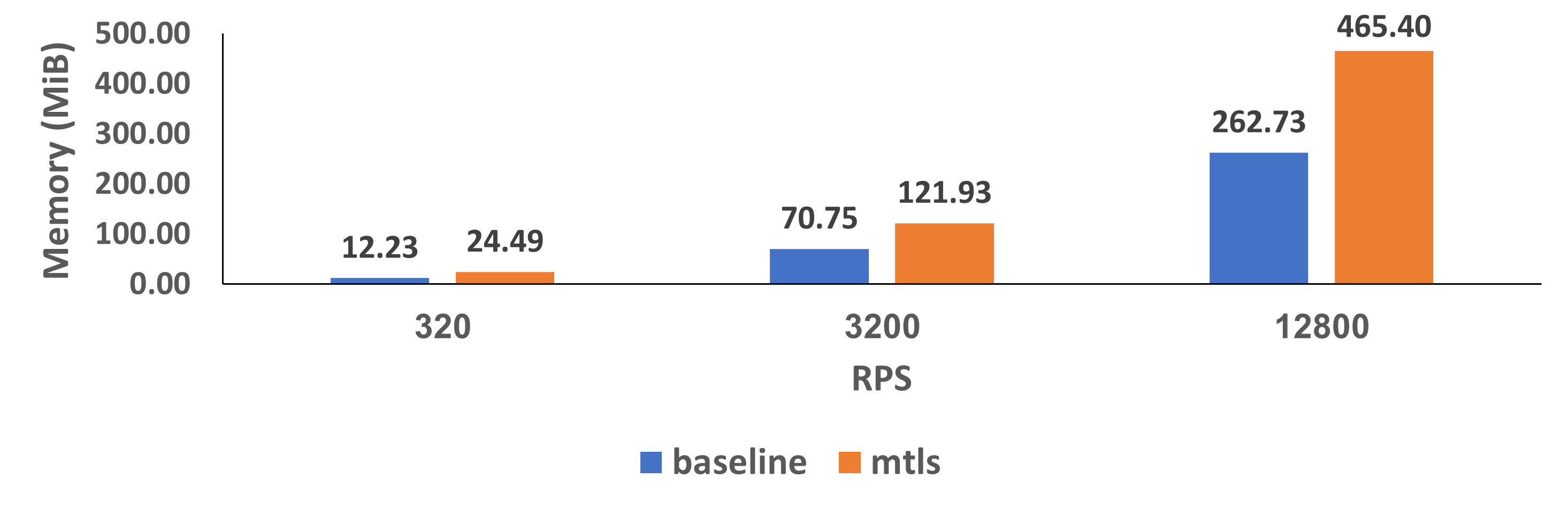}
      \caption{Memory Usage}
      \label{fig:baseline-tests-resources-memory-server}
   \end{subfigure}
   
   \caption{Baseline Resource Tests - Server}
   \label{fig:baseline-tests-resources-server}
\end{figure}

\begin{figure}[h]
       \centering
       \includegraphics[width=\linewidth]{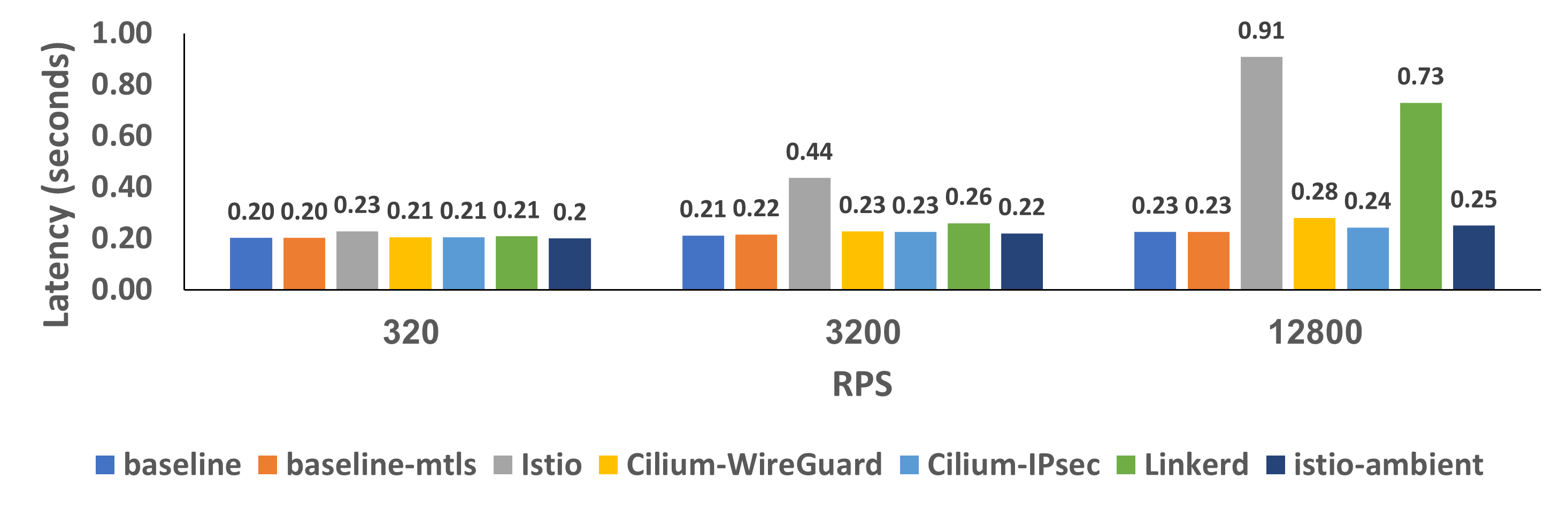}
       \caption{P50 Latency as function of load}
       \label{fig:sm-tests-latency-50}
   \end{figure}

\begin{figure}[h]
   \centering
   \begin{subfigure}[b]{\linewidth}
      \centering
      \includegraphics[width=0.95\linewidth]{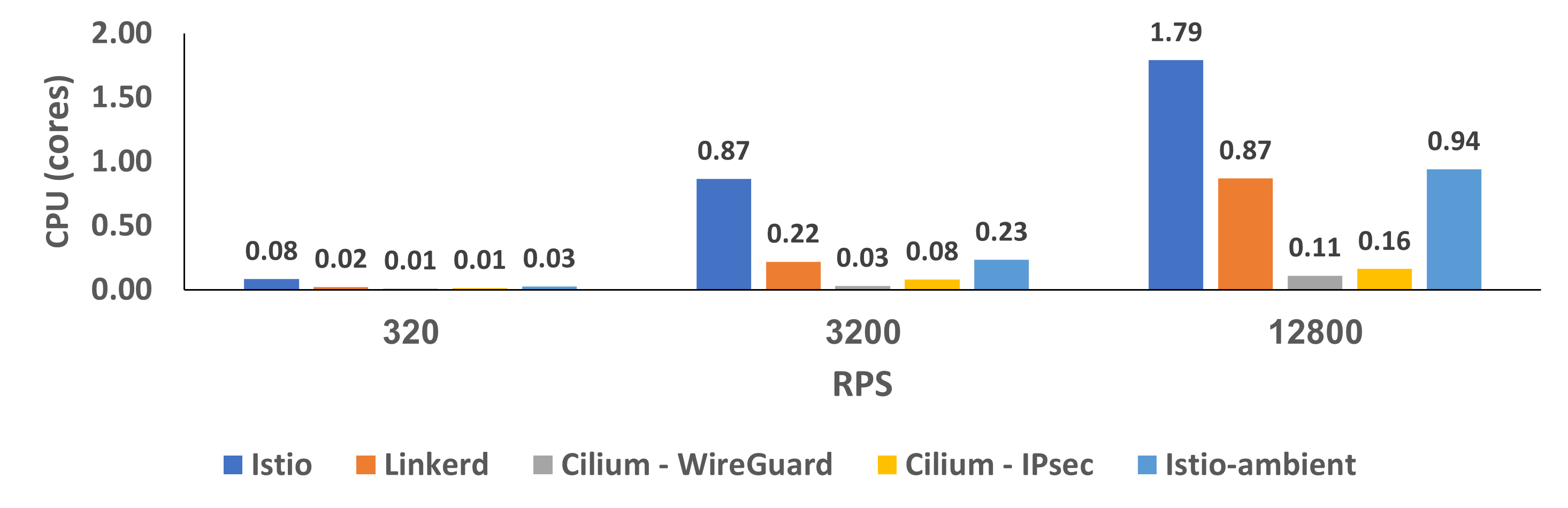}
      \caption{CPU usage}
      \label{fig:sm-tests-cpu-server}
   \end{subfigure}

   \begin{subfigure}[b]{\linewidth}
      \centering
      \includegraphics[width=0.95\linewidth]{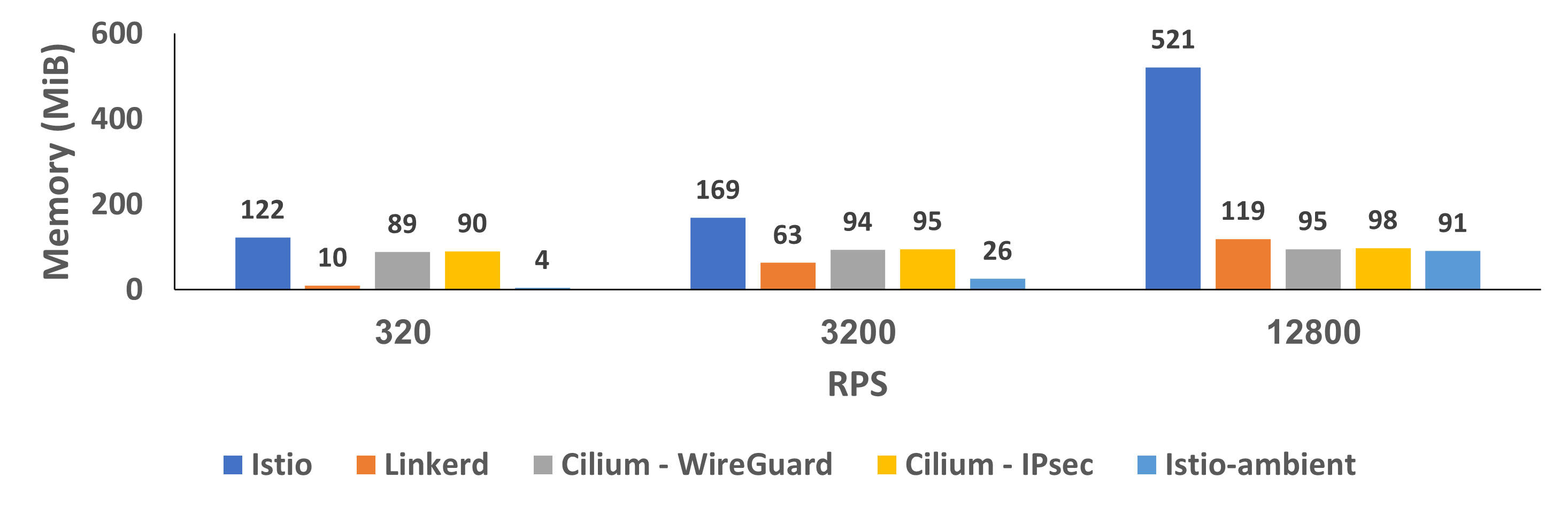}
      \caption{Memory usage}
      \label{fig:sm-tests-memory-server}
   \end{subfigure}

   \caption{Service Mesh's proxies CPU and Memory Usage - Server}
   \label{fig:sm-tests-resource-server}
\end{figure}

\subsection{Configuration of Service Meshes}
\label{sec:config-of-service-mesh}
The service meshes evaluated in this study are designed for Kubernetes environments. Their behavior can be configured through installation arguments, modification of configuration files, or custom resources. Below, we describe the configurations applied to each service mesh to prepare the environment for performance testing.d

\subsubsection{Automatic Sidecar Injection}
Service meshes that have the sidecar pattern approach, have an automatic mechanism for injecting the sidecar proxy into the pods in the service mesh.
In Istio, setting the label \texttt{istio-injection=enabled} on a namespace or \texttt{sidecar.istio.io/inject=true} on a pod triggers injection. Linkerd uses the annotation \texttt{linkerd.io/inject=enabled}.

\subsubsection{Enable or Disable mTLS Protocol}
\label{sec:enable-mtls}
We configured each service mesh to enable or disable mutual TLS (mTLS) as required.
\textbf{Istio}: Uses the \texttt{PeerAuthentication} resource to define traffic encryption modes. \texttt{STRICT} and \texttt{DISABLE} were used to enforce or disable mTLS respectively.
\textbf{Linkerd}: Enables mTLS by default for all TCP traffic between meshed pods. To enforce strict mTLS, the annotation \path{config.linkerd.io/default-inbound-policy=all-authenticated} was used.
\textbf{Linkerd}: We enabled mTLS support in Cilium during installation by adding the flags \path{--set encryption.enabled=true} and \path{--set encryption.type={wireguard, ipsec}} to the \texttt{cilium install} command. For identity management, we deployed the SPIRE server by including \path{--set authentication.mutual.spire.enabled=true} and \path{--set authentication.mutual.spire.install.enabled=true}. Post-installation, we enforced mTLS by adding the attribute \path{authentication.mode=required} to a \textit{CiliumNetworkPolicy} custom resource.

\subsubsection{Istio Explicit Protocol Selection}
\label{sec:istio-protocol-selection}
Istio supports proxying TCP traffic and can automatically detect protocols like HTTP. Disabling protocol detection treats traffic as plain TCP, disabling L7 features.

\subsubsection{Disable Metrics Collection in Istio}
\label{sec:disable-metrics-istio}
We evaluated the overhead introduced by Istio's metrics collection by disabling this feature during installation. We added the argument \texttt{--set values.telemetry.enabled=false} to the \texttt{istioctl install} command.

\subsubsection{Modify Container's CPU and Memory Limits}
To ensure uninterrupted container performance, no CPU or memory limits were set. This configuration allowed the containers to utilize available resources without restriction.

\end{document}